\documentclass[preprint]{elsarticle}

\usepackage{lineno,hyperref}
\usepackage{amsmath}
\usepackage[font=footnotesize,labelfont=bf]{caption}
\usepackage[font=footnotesize,labelfont=bf]{subcaption}

\modulolinenumbers[5]

\journal{Journal of Fusion Engineering and Design}

\begin{document}

\begin{frontmatter}

\title{Robust Vision Using Retro Reflective Markers for Remote Handling in ITER}

%%\author{Elsevier\fnref{myfootnote}}
%%\address{Radarweg 29, Amsterdam}
%%\fntext[myfootnote]{Since 1880.}

\author[mymainaddress]{Laura Gon\c{c}alves Ribeiro \corref{mycorrespondingauthor}}

\cortext[mycorrespondingauthor]{Corresponding author}
\ead{laura.goncalvesribeiro@tuni.fi}

\author[mymainaddress]{Olli J. Suominen}
\author[mymainaddress]{Sari Peltonen}
\author[mysecondaryaddress]{Emilio Ruiz Morales}
\author[mymainaddress]{Atanas Gotchev}
\address[mymainaddress]{Faculty of Information Technology and Communication, Tampere University, 33720 Tampere, Finland}
\address[mysecondaryaddress]{Fusion for Energy (F4E), ITER Delivery Department, Remote
Handling Project 
Team, 08019 Barcelona, Spain}
%\fntext[firstfoot]{These authors contribute equally to the work}

\begin{abstract}
ITER$'$s working environment is characterized by extreme conditions, that deem maintenance and inspection tasks to be carried out through remote handling. 
3D Node is a hardware/software module that extracts critical information from the remote environment during fine alignment tasks using an eye-in-hand camera system and updates the models behind the virtual reality-based remote handling platform.
In this work we develop a retro-reflective marker-based version of 3D Node that estimates the pose of a planar target, the knuckle of the divertor cassette locking system, using the markers attached to its surface.
We demonstrate a pin-tool insertion task using these methods. Results show that our approach works reliably with a single low-resolution camera and outperforms the previously researched stereo depth estimation based approaches.  
We conclude that retro-reflective marker-based tracking has the potential to be a key enabler for remote handling operations in ITER.

\end{abstract}
\begin{keyword}
\texttt ITER \sep Remote Handling\sep Retro-Reflective Markers \sep Eye-in-hand \sep Pose Estimation
\MSC[2010] 00-01\sep  99-00
\end{keyword}

\end{frontmatter}

%\linenumbers

\section{Introduction}
ITER is a worldwide research undertaking that aims at demonstrating the feasibility of fusion as a large-scale source of energy. ITER's experimental machine, the tokamak, has, at its core, a vacuum chamber where fusion occurs. Below the vacuum chamber is the divertor that extracts the heat and ash produced by the fusion reaction and protects the surrounding walls. During the lifetime of the reactor, the activated components of the divertor have to be changed, inspected and repaired.
Since operator access is restricted in the activated areas, maintenance is performed through remote handling (RH).  In particular, the 54 cassette assemblies that comprise the divertor need to be installed and replaced by a remotely controlled robot manipulator that is attached to and carried in by a transporter (cask). After removal, the divertor cassettes are placed in the transporter and taken to another facility, the Hot-Cell. To remove and replace cassettes, it is necessary to operate their locking mechanism, the knuckle, using several tools: a wrench tool, a pin tool, and a jack \cite{VTT2}.The tools must be driven into the respective slots in the surface of the knuckle with narrow gaps and tight tolerances.

ITER's RH is based on tele-operation, which requires the operator to be, at every stage, in the loop and fully in control. In order to perform appropriate actions, the operator has to be presented with all relevant information that portrays accurately the current status of the system. With this purpose, the environment integrates an eye-in-hand camera system to enable augmenting the existing virtual reality models with updated information of the environment. Due to the high radiation levels in the divertor, conventional cameras cannot be used and studies are ongoing on the development of adequate sensors and optical elements to operate in the ITER environment \cite{goiffon2017total, allanche2017vulnerability}.
As it is not yet clear what will be the exact characteristics of ITER's cameras, we assume they will likely provide a low resolution output and a low signal to noise ratio, that degrades over the lifetime of the camera \cite{king2009augmented}, since those are the characteristics of the current analog radiation tolerant cameras. The poor visibility conditions, associated with the lack of some of the natural depth cues, deem a manual alignment based on the camera feeds unreliable. It is necessary to augment the operator's perception during fine alignment tasks \cite{IHA1}, for which a reliable, automatic estimate of the alignment parameters is required.

To address the challenges described herein, the 3D node system was proposed in earlier works \cite{LongChu1,LongChu2,LongChu3}. This solution used the information provided by a stereo camera pair to estimate the geometry of the knuckle through stereo matching. The alignment between the sensed and reference point clouds was obtained using an iterative closest point (ICP) algorithm. Even though the results were promising, the solutions were not robust enough. Some outlier positions were reported, for which the estimated poses were completely in disagreement with the acquired data. Although the reported repeatability of the system is in the $\leq$ 3 mm error range for operating distances of 500 to 1000 mm, the absolute accuracy of the system was not reported. In summary, even though the poses provided by the 3D node seem to be fairly consistent with each other after outlier removal, they do not provide an estimate of what is the agreement between the estimated and the real value of the measures. 
Furthermore, it has been recently assessed that due to space constraints, the system must be simplified to a single camera and operate in the 300-500 mm distance range with an absolute accuracy of $\pm$ 3mm. For the aforementioned reasons, there is a need to develop the 3D node further and find a methodology to estimate its absolute accuracy.

This paper addresses the challenges of the earlier system and proposes a more robust marker-based solution. Markers are often used in photogrammetry for accurate and reliable identification and matching of features \cite{clarke1994analysis}. They provide higher contrast than most naturally occurring structures and are particularly relevant when dealing with large uniform areas and specular surfaces. Specular surfaces differ in appearance with viewpoint and pose a significant challenge to feature detection. The use of retro-reflective markers can further enhance contrast under certain light conditions. These elements have the advantage of offering a particularly high signal to noise ratio, since they reflect most incident light back to the source for a wide range of incident directions. We hypothesize that the introduction of retro-reflective marker tracking in 3D node will greatly increase its robustness and accuracy. 

This paper is organized as follows: Section 2 introduces the architecture of 3D node 2.0.  In Section 3, we describe our marker based approach. In Section 4 we evaluate the performance of our implementation against the former solution. In Section 5 we draw conclusions and outline future development prospects.

\section{System Architecture}

\begin{figure*}
\centering
\includegraphics[width=0.9\linewidth]{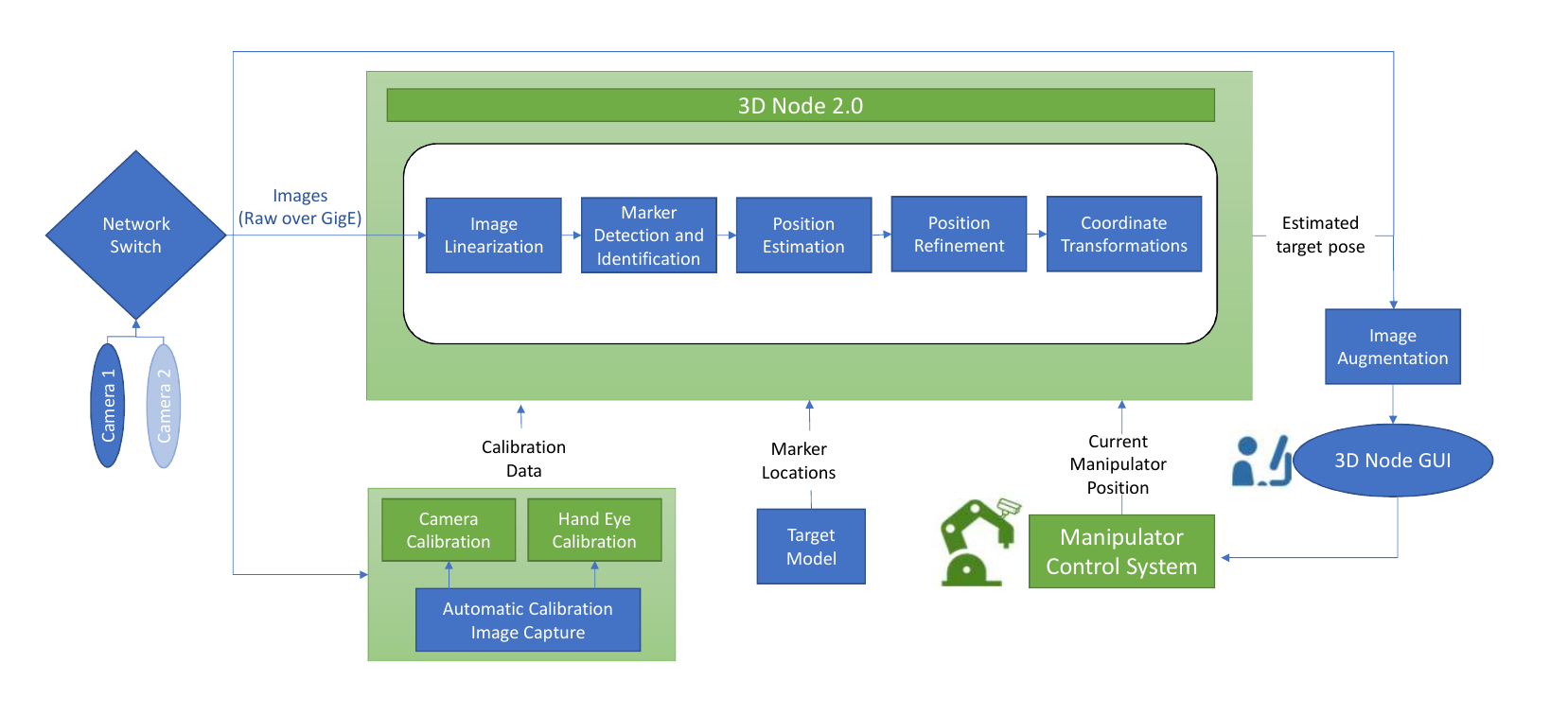}
\caption{Architecture of the 3D node 2.0}
\label{fig1}
\end{figure*}

The architecture of the 3D node has been adapted to include marker-based tracking and is presented in Figure~\ref{fig1}. The new system architecture can obtain an estimate of the alignment using a single camera.
3D node 2.0 receives images in raw format from the camera system through a GigE vision interface. It also receives the information of the pose of the robot end-effector in the coordinate system of the robot base from the manipulator control system (CS).
The calibration module provides the required calibration parameters, which are divided into two groups:  hand-eye calibration and camera system calibration. Hand-eye calibration parameters correspond to the rigid body transformation between the tool center point (TCP) of the robotic manipulator and the optical axis of the camera. Camera calibration parameters can be divided into intrinsic and extrinsic parameters. Intrinsic parameters are the internal parameters of each camera (focal length, principal axis, and lens distortion parameters). Extrinsic parameters correspond to the rigid body transformation between the optical axes of the two cameras. The calibration of the system is performed offline, outside of the operating environment, in the Hot-Cell facility. The calibration module is composed of the automatic calibration image capture stage and estimation stage. In the capturing stage, calibration images and corresponding TCP poses are stored in memory. At estimation stage, the information is loaded and hand-eye and camera calibration are computed independently. 

The pipeline of 3D Node 2.0 is divided into sequential stages: image linearization, marker detection and identification, and pose estimation. The model of the target (in particular, the relative positions of the marker objects) is a required input to the position estimation step. The estimated target pose and the captured images are used to provide feedback to the operator through the Graphical User Interface (GUI). Likewise, the operator provides the manipulation commands to the robot CS through the GUI.

\section{Methodology}

\subsection{Retro-reflective Marker Prototype}

In this section, we describe the developed retro-reflector marker design that complies with the requirements of the ITER operating environment. These requirements can be summarized as follows:
\begin{itemize}
\item Operating temperature up to 200 degrees Celsius.
\item Acceptance for vacuum quality classification 1A and 1B.
\item Tolerant to the radiation in the environment. Furthermore, the radiation in the environment should not alter the properties of the material in such a way that compromises its intended functionality. 
\item Size of the marker should be sufficiently small, such that mounting multiple markers into the cassette is feasible. Moreover, the maximum allowed embedment depth of markers into the cassette is 10 mm.
\end{itemize}
The operating environment is also characterised by the presence of dust. However, dust is not expected to settle significantly on the observed surfaces or camera lens, since these surfaces are mostly oriented vertically. Therefore, we do not consider in this work dust as one of the environment requirements.

In our market study, we found no commercially-available marker that satisfied these requirements and had a suitable performance. Marker performance was evaluated empirically in a practical test setup similar to what would be used in the final version of the system. 
As an alternative approach, we develop a custom made marker object, consisting of an array of KU1 quartz glass beads held by a stainless steel structure. Stainless steel is the main material component in ITER and satisfies all requirements and KU1 quartz glass (fused silica) is approved as a material for all vacuum classes. Moreover, according to the study in \cite{garcia2020}, gamma radiation should not induce significant changes to the refractive index of KU1 glass in the relevant wavelengths (380-700 nm). The reported changes do not influence our application since our measurements are solely intensity-based and are not affected by minor intensity loss.

\begin{figure}
\includegraphics[width=0.48\linewidth]{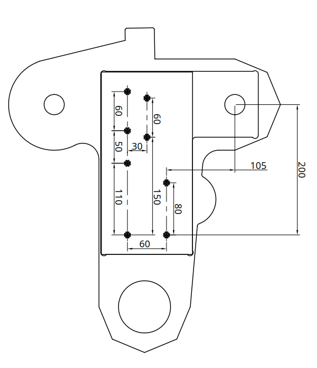}
\hspace*{\fill}
\includegraphics[width=0.42\linewidth]{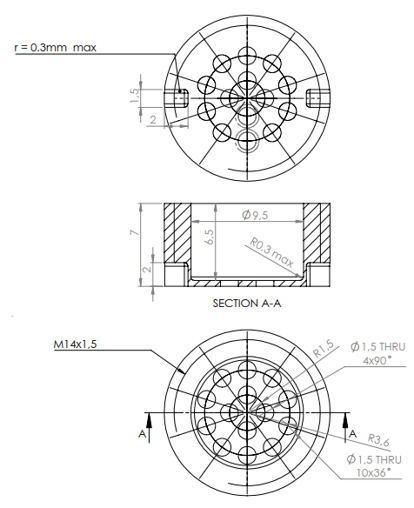}
\caption{Model of target prototype}
\label{fig2}
\end{figure}

The supporting metal structure is a hollow cylinder with 7mm depth, 14 mm diameter and M14 threads on the outside surface (Figure~\ref{fig2}~b). The front surface is a sieve-like structure with 12 holes with 1.5 mm diameter. This design provides attachment and protects the glass beads while simultaneously providing a means of attaching the marker to the cassette. Eight corresponding M14 threads have been incorporated in the cassette design and their relative positions can be seen in (Figure~\ref{fig2}~a). The unified outer shape allows the necessary mounting holes in the cassette to be pre-determined, while allowing future marker development and improvement. Furthermore, the proposed design solution makes it possible to easily change the markers in case any of them sustain damage during operation.

In this section, we have demonstrated that it is possible to manufacture a working retro-reflective marker using ITER compliant materials. 
The described marker was shown empirically to work for the intended purpose, i.e. the contrast between the marker and the surroundings was considered satisfactory in the conditions of the test setup. However, a more dedicated study would be required to maximize the brilliancy of the markers and contrast in the images.

\subsection{3D Node 2.0 Pipeline}

The pipeline of 3D node 2.0 starts with the compensation for lens distortion of the captured images. At the undistortion stage, camera calibration parameters provided by the calibration module are used to correct the lens's optical distortions. 

The following stage in the pipeline is marker detection. Image pixels are segmented into the categories of \textit{marker} or \textit{not a marker} by adaptative thresholding. The applied thresholding method~\cite{bradley2007} can compensate for some illumination variations across the image by comparing each pixel to the values of its neighbours. Due to the specular nature of the knuckle's surfaces, some background regions are classified as \textit{marker} at this stage, a problem that is addressed later by morphological filtering.
The discernible small white elements of each marker array are merged into a single elliptic shape by image processing operations. We apply morphological closing and flood-fill operations~\cite{soille2013morphological} consecutively, followed by determination of the binary convex hull image. The blobs in the resulting image are identified by connected component labelling~\cite{haralick1992computer}. False blobs are excluded based on their geometric properties (such as area and eccentricity). We find the coordinates of the centroid of each detected blob. Assuming that the target and camera have been roughly aligned by the operator allows us to make the approximation that the centroid of the detected blobs roughly corresponds to the centre of the original circles.

Next we must establish a one-to-one correspondence between the known relative marker 
positions in the world coordinate frame and the detected image coordinates. Markers are ordered and identified according to their relative Euclidean distances. This methods works only for small rotations of the target, since the distance between image points is not a perspective invariant.

Once the planar point correspondences have been established, we  use a homography decomposition approach~\cite{hartley2003} to determine the initial estimate of the rigid body transformation between the camera and the target. 
At this stage, we check the reprojection error provided by the current estimate. If it is higher than a few pixels, we go back to the marker identification and ordering stage and ask the operator to intervene in the selection of the thresholding and filtering parameters.
The initial pose estimate is further optimized through minimization of the reprojection error using the Levenberg-Marquardt algorithm~\cite{levenberg1944method, marquardt1963algorithm}. In the case of stereo cameras, we take as initial estimate the mean value of the estimates provided by the two cameras and optimize the results based on the point correspondences from both images.

The possibility of optimizing over several images taken by the same camera at different poses is studied in another work~\cite{ihtishams}. In this work two multiview estimation methods have been considered: (1) using the poses reported by the robot's CS and (2) not using those. Further on, we provide a comparison with those methods, refered to as "Multiview 1" and "Multiview 2", respectively.

The calculated target pose in the coordinate system of the camera must be converted to valid commands in the coordinate system of the robotic arm. For this, an estimation of the hand-eye transformation is required. A detailed survey of hand-eye calibration methods is presented in~\cite{Ali2019}. From the surveyed methods we have utilized here the method of Shah~\cite{shah2013solving}.

\section{Experiments And Results}
\subsection{Equipment and Setup}
\begin{figure}
\centering
\includegraphics[width=0.9\linewidth]{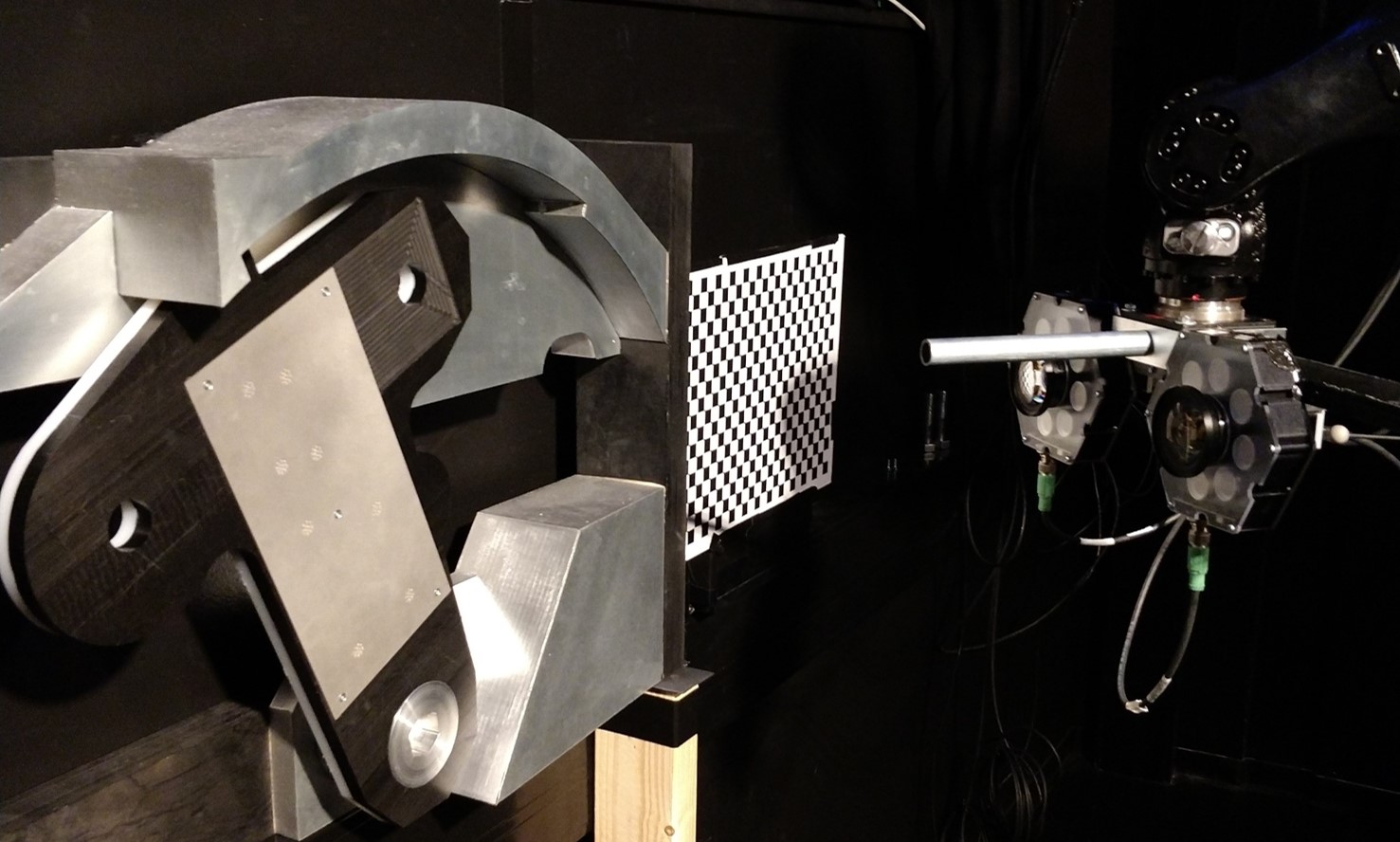}
\caption{Illustration of the experimental setup. On the left side, the mock-up target object. On the right side, the camera system, ring lamps and mock-up pin-tool, attached to the end effector of the KUKA robot manipulator.}
\label{fig3}
\end{figure}

The KUKA KR 16 L6-2 industrial robot is used in our experimental setup to simulate ITER's robotic arm, which is still under development. In our laboratory, the robot is fixed to a static platform, contrary to the ITER design, in which the arm is fixed to a structure that moves on rails. KUKA has a lower payload capacity, requiring the use of lighter weight mock-up versions of the tools. A version of the pin-tool was created for demonstration purposes, where a pipe segment with the corresponding diameter represents the pin. On the right side of Figure~\ref{fig3} we show the end-effector of the manipulator and the prototype pin-tool.

As a target, we use a 1:1 scale 3D printed replica of the Cassette Locking System (CLS). Since the conditions in the laboratory environment do not pose the same strict material restrictions as the application, a simplified version of the retro reflective markers and their holders was made for testing. In the simplified version, the bead holding arrays are laser cut into a 1.0 mm thick sheet of metal, as shown in Figure~\ref{fig2}~a, making up a single holding plate. Prototype beads are made of borosilicate glass. This material does not comply with the ITER environment requirements, but is used to simulate the optical behaviour of fused silica at a cheaper cost.
The metal sheet is attached to a back plate made of polycarbonate that simulates the knuckle geometry and is fitted onto the CLS replica. An image of the prototype is shown on the left side of Figure~\ref{fig3}.

A stereo pair of industrial Basler acA1920-48gc cameras was rigidly attached to KUKA's end effector, as shown on the right side of Figure~\ref{fig3}. The cameras have a native resolution of 1920x1200 pixels with an RGB Bayer filter to produce colour images. To emulate the use of radiation tolerant cameras, we convert the images to grayscale and downsample the resolution by a factor of three to 640x400 pixels. 
The cameras are rigidly attached to an aluminium holding piece with a baseline of 200 mm. The baseline is dictated by the size of the lighting elements and the construction of the mock-up tool holder. The design of the mounting is not indicative of the one to be used in the ITER environment, and a different baseline can be designed as needed. The lenses are high quality, low distortion wide angle lenses with a nominal 6 mm focal length. With these particular cameras, the lenses produce an opening angle of roughly 82 degrees. Two Smart Vision RC130 industrial ring lights were fitted to the camera pair, as closely as possible to the camera lenses. These lamps create a relatively uniform light pattern that covers the entire target for an acceptable variety of camera poses.

\subsection{Accuracy and Precision of 3D Node 2.0}

The ground-truth values for assessing the accuracy of the system are obtained by manually positioning the robot's end effector at the target origin. This allows us to calculate, for each test pose, the ground-truth transformation between the camera and the target, based on the values reported by the robot's CS.

Given a rigid body transformation, 
$
T= 
\begin{bmatrix}
R & t\\
0 & 1
\end{bmatrix}
$
, and the corresponding ground-truth transformation,
$
T^{GT}= 
\begin{bmatrix}
R^{GT} & t^{GT}\\
0 & 1
\end{bmatrix}
$
, where R represents the 3X3 rotation matrix and t represents the 3X1 translation vector,
we calculate the translation error vector as

\begin{equation}
t^{e}=t - t^{GT}
\label{eq2}
\end{equation}

We calculate absolute translation error $(e_t)$ in millimetres and the absolute rotation error $(e_R)$ in degrees as follows

\begin{equation}
e_t=\sqrt{(t^{e}_x)^2 + (t^{e}_y)^2}
\label{eq2}
\end{equation}

\begin{equation}
e_{R}= |angle(R)-angle(R^{GT})|
\label{eq3}
\end{equation}

As seen in Equation~\ref{eq2}, we evaluate translation error as the Euclidean distance in each of the two axes that are parallel to the target plane. We disregard the depth axis, since it is not relevant for the alignment task.

In Equation~\ref{eq3}, angle() represents the conversion from a rotation matrix to axis-angle representation and consideration of only the attitude quaternion.

We evaluate performance over a set of images captured from 20 random poses within the following limits:
\begin{itemize}
\item Distance of 300 to 500 mm from the target.
\item Horizontal displacement from the centre of the target smaller than 50 mm.
\item Vertical displacement from the centre of the target smaller than 100 mm.
\item Rotations in relation to the aligned pose smaller than 10 degrees.
\item The target is fully visible by both cameras.
\end{itemize}

\begin{figure}
\subcaptionbox{\label{subfig-1:platzhalter}}{%
      \includegraphics[width=0.45\linewidth]{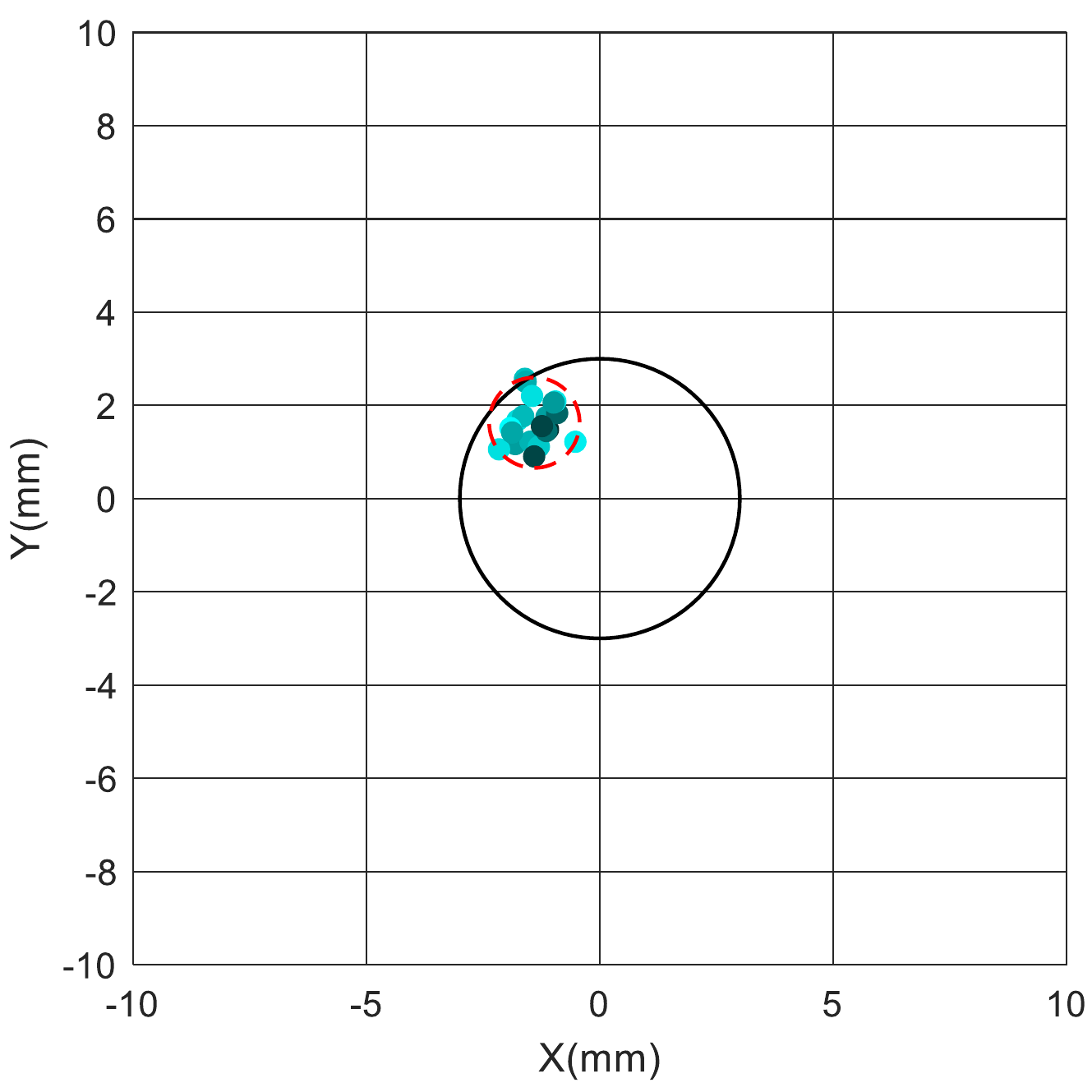}%
    }
\hspace*{\fill}
\subcaptionbox{\label{subfig-2:platzhalter}}{%
      \includegraphics[width=0.45\linewidth]{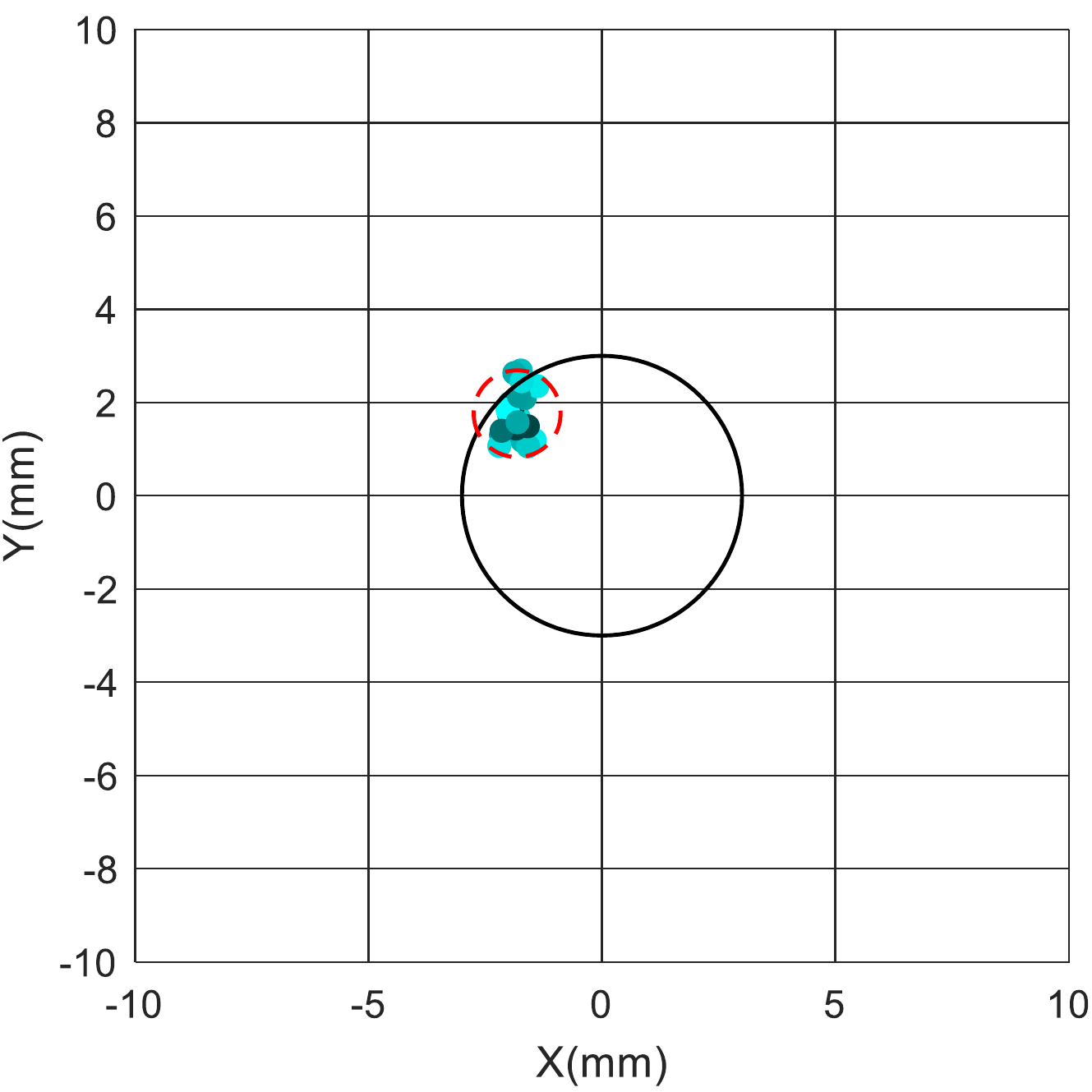}%
    }
\caption{Representation of absolute error in each of the X and Y axis for the estimation provided by left most camera (\textbf{a}) and stereo (\textbf{b}). Each colour point represents a different pose of the camera in relation to the target.}
\label{fig4}
\end{figure}

The x and y components of the error vector ($t^e$) are shown in Figure~\ref{fig4} for the proposed marker-based monocular and stereoscopic approaches. In our visual representations, the red dashed circle is centred in the mean error value and its radius corresponds to the maximum distance between the observations and the center point. The black circle is centred in (0,0) and its diameter corresponds to the maximum allowed error. The requirements of the application have been fulfilled when the dashed red circle is contained within the black circle. 

Analysing the results in Figure~\ref{fig4}, we see that the application requirements have been fulfilled and errors are distributed within a fairly narrow region even though we tend to consistently misestimate translation by about 2 millimeters in both the X and Y directions. 

\subsection{Markerless Approach}

In the earlier version of 3D Node~\cite{LongChu2} the pose estimation problem was tackled by fitting a stereoscopic depth estimate to a known CAD model using an ICP algorithm. To compare with this approach, we captured the same sequence of robot poses with the same scene object, but with increased lighting to enable the stereo cameras to see the objects. We give the ground truth pose with some added noise as an initial estimate to the algorithm to make sure it converges on the correct solution, so we can reliably measure its accuracy without concern about parameter tuning. 

\begin{figure}
\subcaptionbox{\label{subfig-1:platzhalter}}{%
      \includegraphics[width=0.45\linewidth]{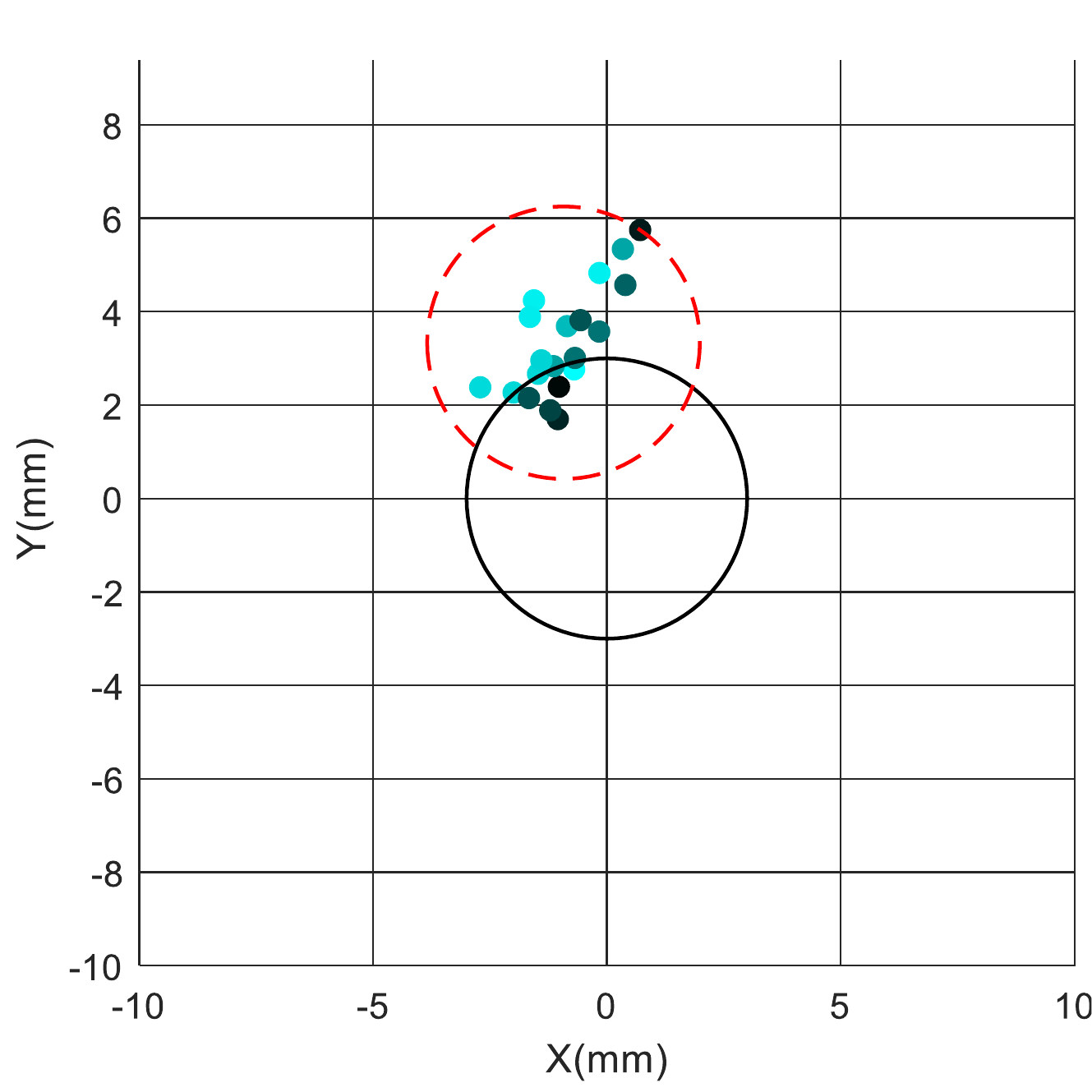}%
    }
\hspace*{\fill}
\subcaptionbox{\label{subfig-2:platzhalter}}{%
      \includegraphics[width=0.45\linewidth]{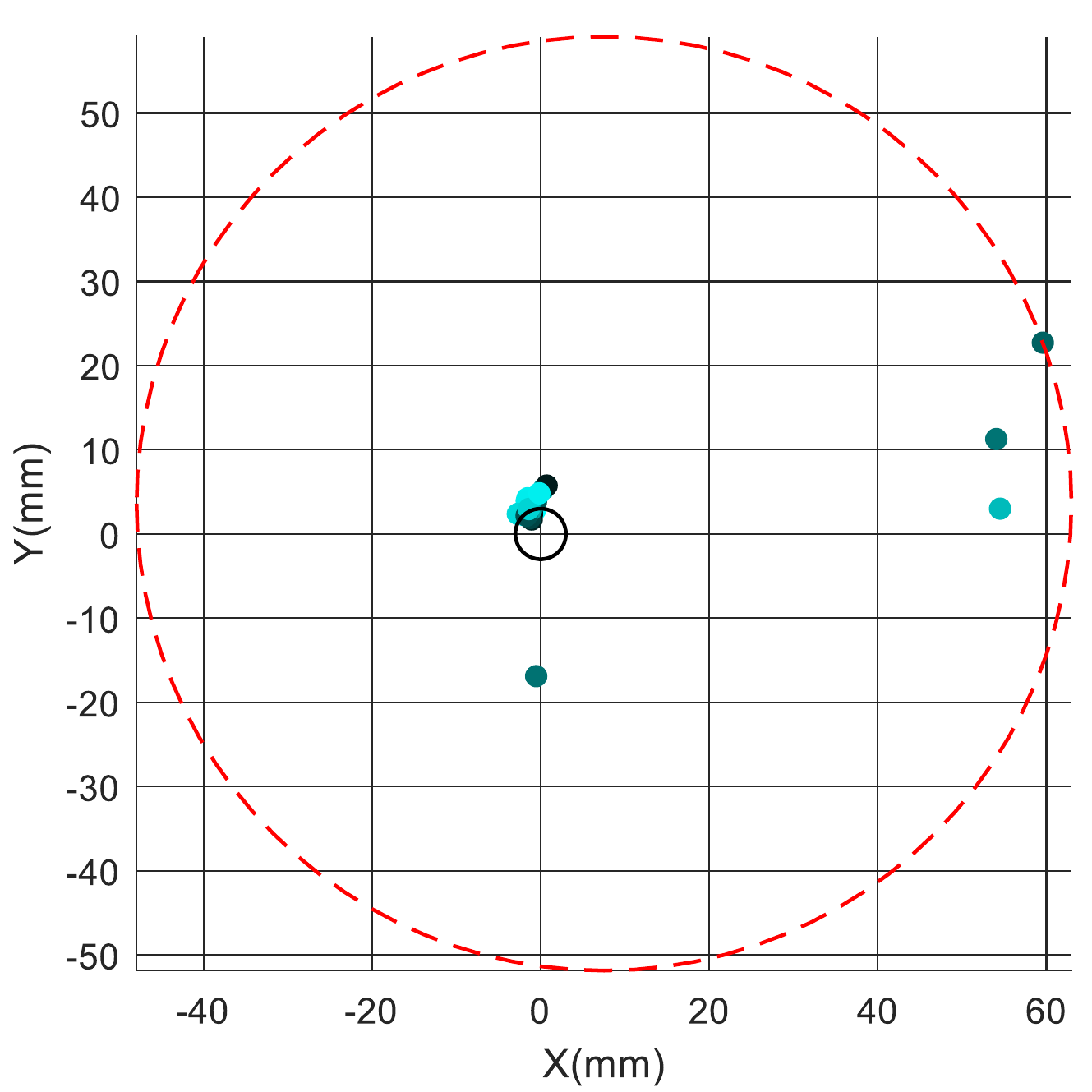}%
    }
\caption{Representation of absolute error in each of the X and Y axis for the estimation provided by the stereo ICP based method from earlier work, with different amounts of initial pose deviation from ground truth. Offset values of $|\Delta a|=$5 deg, $|\Delta t|=$ 25 mm (a). Offset values of $|\Delta a|=$ 10 deg, $|\Delta t|=$ 50mm (b). Each colour point represents a different pose of the camera in relation to the target.}
\label{fig5}
\end{figure}

We observe that with small initial offsets, it converges closely to the expected values (Figure~\ref{fig5}, left side) though it results in a spread of $\pm 3mm$, barely within the specification. With larger amounts of offset (Figure~\ref{fig5}, right side), the convergence deteriorates quickly and no longer ends up with correct estimates.

\subsection{Marker-based Multiview Pose Estimation}

To be able to establish a comparison with the marker-based multiview estimation methods of~\cite{ihtishams} (Multiview 1 and Multiview 2), for each image in the dataset described in section 4.1, we record an extra set of 5 random poses.

\begin{figure}
\subcaptionbox{\label{subfig-1:platzhalter}}{%
      \includegraphics[width=0.45\linewidth]{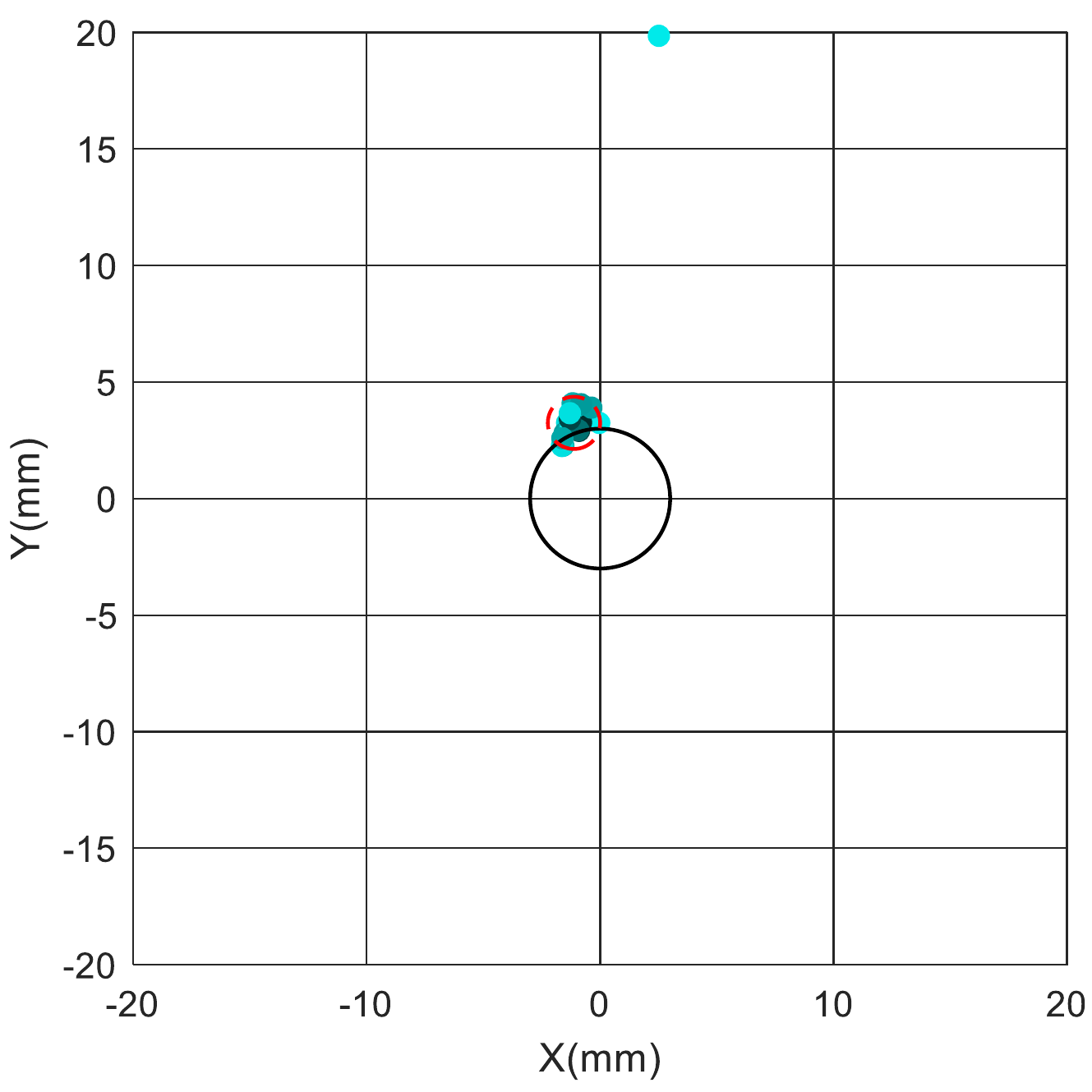}%
    }
\hspace*{\fill}
\subcaptionbox{\label{subfig-2:platzhalter}}{%
      \includegraphics[width=0.45\linewidth]{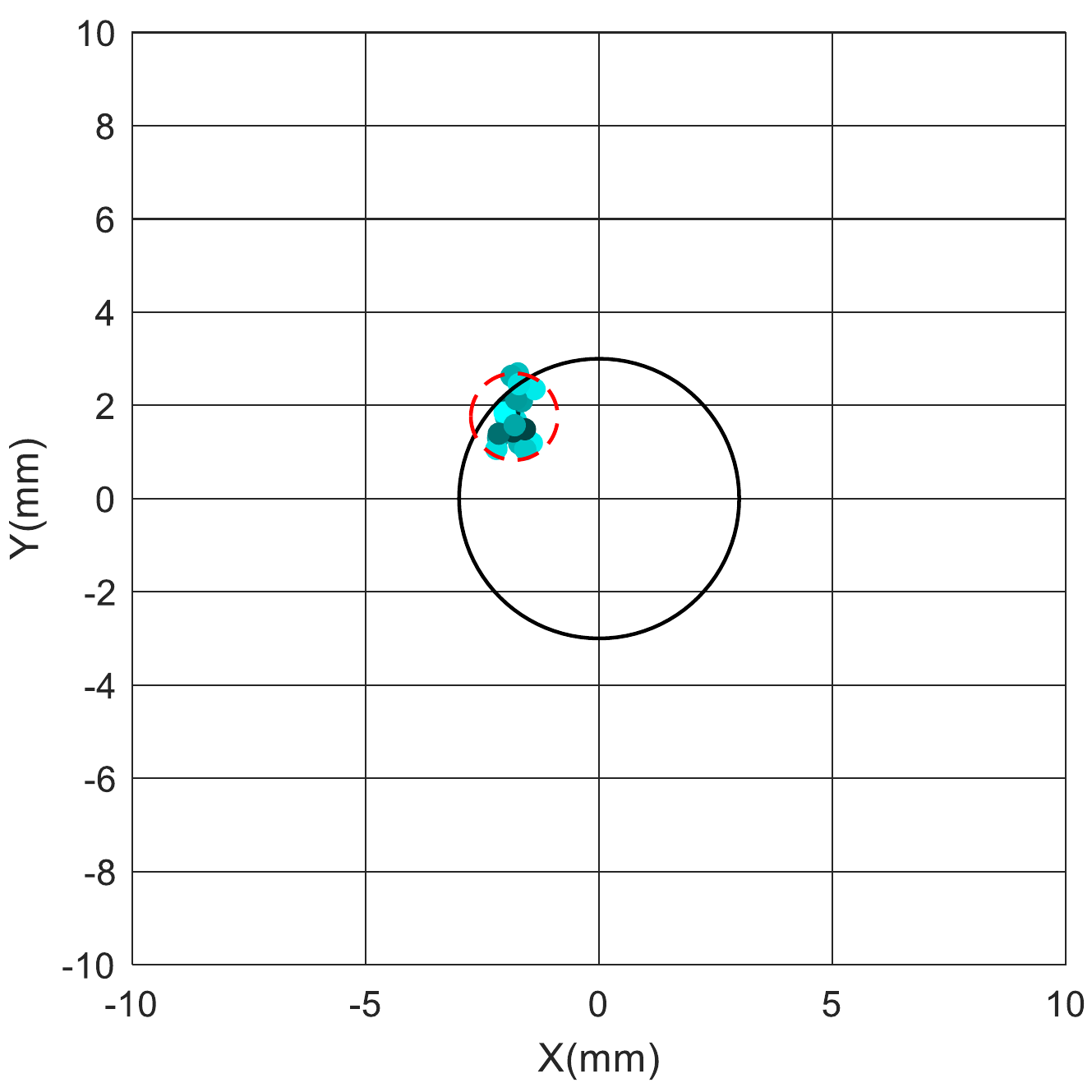}%
    }
\caption{Representation of absolute error in each of the X and Y axis for the estimation provided by the multiview method 1(\textbf{a}) and 2 (\textbf{b}). Each colour point represents a different pose of the camera in relation to the target.}
\label{fig6}
\end{figure}

Results are presented in Figure~\ref{fig6}. We observe that both methods work quite consistently, despite method 1 having one outlier, for which the error is significantly higher.

\subsection{Comparative Analysis}

Comparative results are presented in Tables~\ref{table:allresultsT} and~\ref{table:allresultsR}. In these tables, precision refers to the comparison of a estimated value with the average of estimates, while accuracy refer to the comparison to the ground truth. We present the average and maximum error values among the 20 observations for all evaluated approaches.
To compute the statistics in the tables we use the results of ICP with smaller initial offsets (corresponding to the left side of Figure~\ref{fig5}) and exclude the outlier seen in the left side of Figure~\ref{fig6}.

\begin{table*}[ht]
\caption{Average and Lowest Accuracy and Precision of Estimated Translations.} % title of Table
\centering % used for centering table
\begin{tabular}{ *{5}{c} }

     & \multicolumn{2}{c}{Accuracy} & \multicolumn{2}{c}{Precision}  \\
     $e_t(mm)$ & Average &  Maximum &  Average &  Maximum \\
    \hline
    Monoscopic & \textbf{2.1944} & 3.0295 & 0.5597 & 0.9691 \\
    \hline
    Stereoscopic & 2.2924 &  \textbf{2.9813} &  0.5033 & 0.9707\\
    \hline
    Stereoscopic ICP~\cite{LongChu2} & 3.6017 & 5.7915 & 1.2287 & 2.9162  \\
    \hline
    Multiview 1 & 3.4751 & 4.2541 & 0.6178 & 1.1204 \\
    \hline
    Multiview 2  & 2.5606  & 3.2318    &   \textbf{0.4797}    & \textbf{ 0.9304}  \\
    \hline
\end{tabular}
\label{table:allresultsT} % is used to refer this table in the text
\end{table*}

\begin{table*}[ht]
\caption{Average and Lowest Accuracy and Precision of Estimated Rotations.} % title of Table
\centering % used for centering table
\begin{tabular}{ *{6}{c} }

     & \multicolumn{2}{c}{Accuracy} & \multicolumn{2}{c}{Precision}  \\
     $e_R(^{\circ})$ & Average &  Maximum &  Average &  Maximum \\
    \hline
    Monoscopic & 0.4105 & 1.6886 & 0.2421 & 1.2781 \\
    \hline
    Stereoscopic & \textbf{0.3291} & \textbf{0.7950} & \textbf{0.1563} & \textbf{0.4659}  \\
    \hline
    Stereoscopic ICP~\cite{LongChu2} & 0.5367 & 1.5510 & 0.3098 & 1.0143 \\
    \hline
    Multiview 1 & 0.4735 &  1.0074 &  0.2292 & 0.5339 \\
    \hline
    Multiview 2  &  0.5381 &   1.2164  &   0.2590 &   0.6783\\
    \hline
\end{tabular}
\label{table:allresultsR} % is used to refer this table in the text
\end{table*}

From the analysis of the information in the tables, we see that both marker-based methods, either with one or two cameras, outperform the earlier version of the system. The error of both single and stereo marker-based methods is about three times lower than ICP's. The earlier version has a slightly lower maximum angular error compared to the single camera marker, but only to the scale of 0.1-0.2 degrees. 

In summary, the new approach produces much more consistent results than the previous one. Though not a part of this numerical analysis, the new marker based method is also much more robust in finding the correct alignment, and does not in fact require any initial estimates to function properly. 

Stereoscopic estimation with markers performs slightly better on the metrics of orientation estimation, but does not perform overall significantly better than monocular and it actually loses to monocular in average translation accuracy. 

Regarding the approaches using multiple shots taken with the same camera, both approaches outperform monoscopic estimation in maximum angular error. Method 1 estimates rotation consistently better than Method 2. However, Method 1 shows the highest precision of all methods in estimating translation, although the difference is not very significant.

We believe that the offset of about 2 mm in the x and y axes that can be seen in the results is due to errors in the poses reported by the robotic arm's CS. This is likely due to the fact that the robot was uncalibrated. These hypotheses were not confirmed, and further study is required.

\subsection{Pin-tool Insertion Demonstration }

\begin{figure}
\hspace*{\fill}
\subcaptionbox{\label{subfig-1:platzhalter}}{%
      \includegraphics[width=0.9\linewidth]{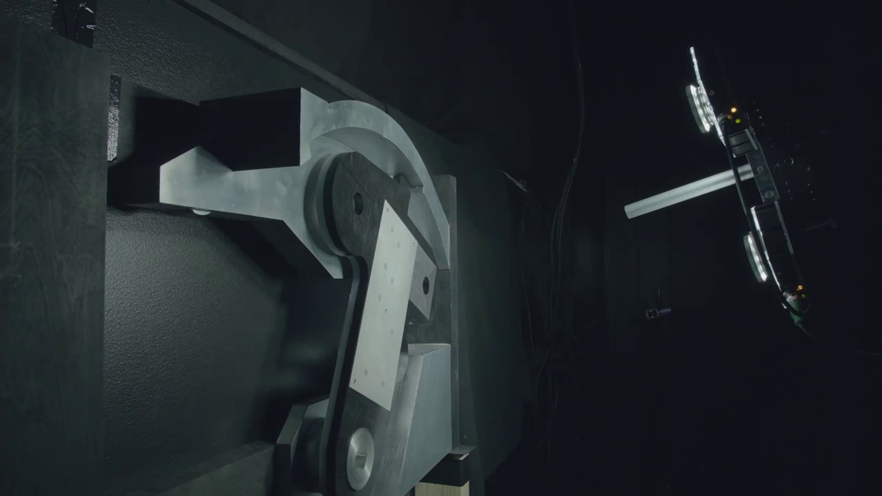}}
\hspace*{\fill}
\subcaptionbox{\label{subfig-2:platzhalter}}{%
      \includegraphics[width=0.9\linewidth]{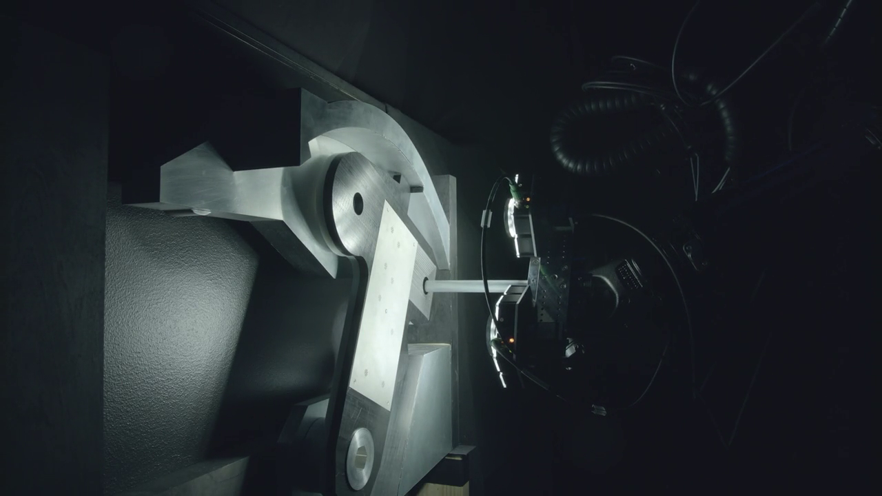}}
\hspace*{\fill}
\subcaptionbox{\label{subfig-3:platzhalter}}{%
      \includegraphics[width=0.9\linewidth]{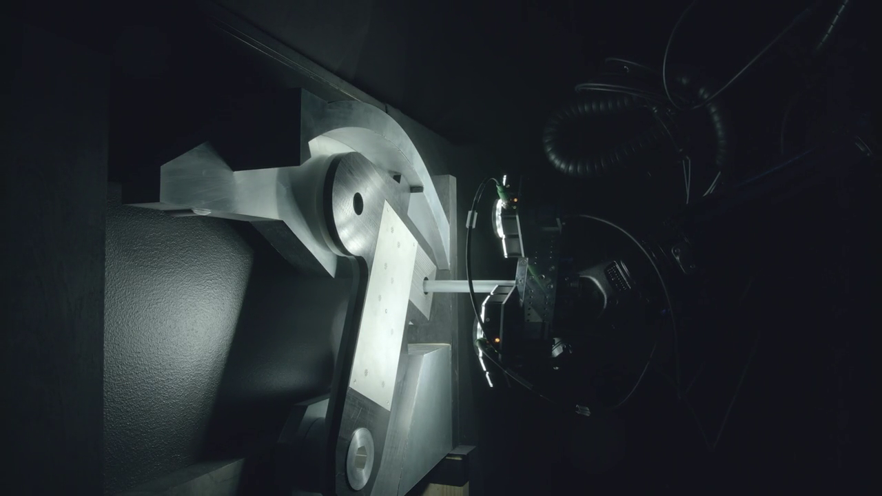}}
\caption[]{Demonstration of pin-tool insertion. Random initial pose (a), aligned pose (b), final pose (c). These images are screenshots from the demonstration video that can be found in \cite{demoVideo}}
\label{fig7}
\end{figure}

We set up a proof of concept demonstration of the insertion of the pin-tool in the respective slot in the cassette surface based on the estimate provided by the single-camera method. In the demonstration, the manipulator moves autonomously to the aligned pose, based on the estimate provided by 3D Node 2.0 (Figure~\ref{fig7} b). The aligned pose is such that the tool and target planes are aligned, but the tool is still at a distance from the target. After this movement is completed, the operator confirms the approximation to the target and the tool is inserted (Figure~\ref{fig7} c). 
The demonstration is performed for the same set of 20 randomly chosen poses as described above. The insertion of the tool was successful for all the poses.

\section{Conclusions and Future Work}

In this work, we have demonstrated a retro-reflective marker design that complies with both vacuum class and radiation requirements of ITER's operating environment. We showed that the proposed retro-reflective marker-based implementation of 3D node works more reliably and provides more accurate results than its earlier version. 
The overall performance of 3D node 2.0 is within the required tolerance ($<\pm3$ mm) both in the single camera and stereo camera modes. In our results, for the tested viewing range (300-500 mm) and baseline value (200 mm), there is no major difference in performance between the monocular, stereoscopic and multi-view approaches. We can conclude that the proposed marker-based approach in any of its modes shows potential to be a key enabler for RH in ITER.

Further study is needed on the refinement and optimization of all elements in the system, in particular: 
(1) optimization of the retro reflective marker design to maximize brilliancy for the observation angle, 
and (2) use of more advanced approaches for marker identification. 
Further work could also quantify the comparative performance of retro-reflectors and standard diffuse markers in the ITER environment, once characteristics of the environment, such as existence of ash, dust or other particles, has been determined. Further work could also address how the calculated alignment (translation vector and rotation matrix) can be optimally presented to the operator and integrated within the existent augmented and virtual reality solutions.

\section*{Acknowledgements}
The work in this paper was funded by the European Union's Horizon 2020 research and innovation program under the Marie Sklodowska Curie grant agreement No 764951, Immersive Visual Technologies for Safety-Critical Applications and by Fusion for Energy (F4E) and Tampere University
under the F4E grant contract F4E-GRT-0901.
The research infrastructure of Centre for Immersive Visual Technologies (CIVIT) at Tampere University provided the robotic manipulator, sensors and laboratory space for conducting the experiments.
 This publication reflects the views only of the authors, and Fusion for Energy cannot be held responsible for any use which may be made of the information contained herein.

\bibliography{mybibfile}

\end{document}